\documentclass[twocolumn,twoside,slac]{revtex4}
\usepackage{graphicx}
\usepackage{fancyhdr}
\begin{document}

\title{An Object-Oriented Minimization Package for HEP}


\author{M. Fischler and D. Sachs}
\affiliation{FNAL, Batavia, IL 60510, USA}

\begin{abstract}
 A portion of the HEP community has perceived the need for a minimization 
 package written in C++ and taking advantage of the Object-Oriented nature
 of that langauge. 
 To be acceptable for HEP, such a package must at least 
 encompass all the capabilities of Minuit. 
 Aside from the slight plus of 
 not relying on outside Fortran compilation, the advantages that a C++ 
 package based on O-O design would confer over the multitude of available 
 C++ Minuit-wrappers include: Easier extensibility to different algorithms 
 and forms opf constraints; and usage modes which would not be available 
 in the global-common-based Minuit design. 
 An example of the latter is a 
 job persuing two ongoing minimization problems simultaneously. 
 We discuss the design and implementation of such a package,
 which extends Minuit only in minor ways but which greatly diminishes the 
 programming effort (if not the algorithm thought) needed to make more 
 significant extensions.
\end{abstract}

\maketitle

\thispagestyle{fancy}


\section{MOTIVATION}

\subsection{Minuit}

For many years, the gold standard in  minimization packages used by 
High Energy Physicists has been the Fortran program  
Minuit~\cite{minuit}.
This code contains algorithms which are well-suited for many HEP uses, 
especially when there is a large data set and minimization is needed to find a
multi-parameter fit.

The principal algorithm, called {\sc migrad}, is a variable metric method.
This exploits the idea that althought a costly second derivative computation
is needed to move to the function minimum, that matrix need not be
computed accurately for early steps (since it will be changing
significantly anyway).
Instead, it can be iteratively improved, 
so that by the time the method gets near
the actual minimum, the matrix is accurate and rapid convergence occurs.

Another prime capability is a set of routines for analyzing the solution:
The {\sc hesse} method of supplying the parameter 
error correlation matrix,
the {\sc minos} method of nonparabolic chi-suqare error intervals, 
and {\sc contour}, which depicts the shape of a 
fixed-sigma error curve around the solution point for functions which may
not be well-described as bi-linear forms.

\subsection{Is a New Minimization Package Needed?}

Minuit, however, is getting a bit stale.  It is written in Fortran, which
is no longer the {\em lingua franca} of computing in 
HEP; it can be awkward to link C++ code with
code from a different language.  
And modern users would like certain advantages relating to treating a
minimization problem and its function, algorithms and domains, 
as objects.  One such advantage is the option of performing
multiple independant minimizations in tandem.
 
The Minuit code is difficult to maintain, as evidenced by the fact that 
there have been no enhancements since a major overhaul done by F. James about
a decade ago, and no major algorithm
improvements for some time before that.
The HEP community knows of promising alternative methods, such as 
FUMILI~\cite{fumili} and the related LEAMAX~\cite{leamax}, and certainly the
field has advanced in these years, with simulated annealing and genetic
algorithms maturing.  The fact that nobody has seen fit to incorporate these 
as options in Minuit can be taken as evidence that such changes
would be hard to make.

Even assuming that Minuit needs no maintenance because 
all bugs have been eliminated and that its capabilities are
forever adequate, what 
happens when all the CERNLIB Fortran modules start becoming awkward
to bring forward?  There is reason to fear that Minuit might 
fall into functional obsolecence.
 
We asked 
a group of HEP physicists who work heavily with C++ computation 
whether it would be a good idea to produce a C++ object-oriented minimization package
with Minuit's capabilities.  
About 40\% reacted positively:  ``it's about time we had this.''
Some of these physicists had
faced a tougher time accomplishing 
an actual project, for lack of a good stand-alone C++ minimizer.

Others reacted with some variant of ``are you
out of your mind?  Sombody must already have done this!''  We will argue
below that nobody actually has delivered an adequate object-oriented 
minimizer with the capabilities required.

\subsection{Other Minimizers}

Most commercial minimization packages fall short in capability
aspects.  In some cases, their algorthms are not as appropriate as {\sc migrad}
for the sort of problems which occur in HEP; more often they lack solution
analysis methods equivalent to {\sc contour} and {\sc minos}.  
Also, non-trivial licensing issues can hinder 
the adoption of commercial codes as {\em de facto} standards in the
HEP community. 

A more promising approach is to translate Minuit directly into C or C++.  
Several analysis packages, including ROOT~\cite{root}, provide fitting in 
this manner.  
Another approach is to place a wrapper around the Minuit 
code, to allow the interface to be expressed in C++, and package
the compiled Fortran code along with the wrapper as an integrated library.
Gemini~\cite{gemini} does this, providing a shell over a choice of Minuit 
or the commercial NAG C minimizer.

Either approach solves the problem of inter-language awkwardness, 
but they do not address  the issue of maintainability and ease of
enhancement.  And fundamentally, a minimization problem still cannot
be treated as a first-class object.

The Root team, 
in particular, has done significant work both in cleaning up the 
{\em f2c} translation of Minuit and in removing deficiencies involving 
multiple minimizations.  But not everybody welcomes the requirement 
of linking to the large body of Root code,
when all that is needed is a minimizer.
  
One other C++ object-oriented minimizer is definitely worth mentioning.
Matthias Winkler, mentored by Fred James at CERN, is working on an
``object-oriented re-implementation of Minuit in C++'' under the auspices of the
SEAL programme.  
Clearly 
two nearly-identical 
good C++ minimizers are not needed.
We have been in contact with Dr. James, and intend to coordinate the two
parallel efforts, hoping to emerge with a single package of superior design
and implementation quality.  

\subsection{Minimizers vs. Fitters}

It is worth saying a few things about the concept of a minimizer as 
opposed to a fitter.  Since most applications of minimization in HEP are
motivated by multi-parameter fits to data, the two can be confused.

Most of the packages which wrap Minuit tend to be fitters, that
is, they present a convenient way to provide a function form and a 
data set, and go about using minimization to optimize the parameters in
a fit to that data.  When the function $f$ is a chi-squared sum, the
resulting minimization procedure is specialized, 
in the sense that the nature of the fitting problem assigns a meaning 
and a natural scale for differences in values of $f(\vec{v})$.   
Algorithms (such as {\sc migrad}) which yield a meaningful 
estimated distance above the true minimum can exploit a natural termination
condition, since the value of $f(\vec{v})$ assumed to be accurate only
to some small fraction of a unit in chi-squared.  And analysis of the 
estimated shape of $f(\vec{v})$ can meaningfully be translated to statistical
statements about correlations among the parameters.

However, the kernel operation of minimizing an arbitrary function is
a valid capability to ask for.   
By design, the C++ minimization package under
development is a minimizer, not a fitter.
One can automate the steps
to turn a fitting task into a minimization problem.
Several others have done this in C++, and though it may
become convenient to package such a fitter 
adaptor with the minimizer,  this adaptor
need not be considered part of the development of the minimizer.

\section{PACKAGE REQUIREMENTS}

The requirements guiding the developement of this package
are driven by the goal of 
acceptance and widespread use in the HEP community.
This will require providing a significant set of capabilities, discussed below.
The package must not to be deficient, with respect to Minuit, 
in any capability. 
Also, attention must be paid to certain performance considerations.
The package must be clean to use and self-contained,
and should not depend on other packages.
And stringent testing will be required, to ensure that the package is robust.
  
At the same time, we wish to  
provide interfaces which will encourage code 
which is clear and readable and which does not deviate from the classical C++ 
coding idioms accepted by top practitioners.

Our goal is to provide a package that can be used for many years.
It is impractical to provide something which 
is ``so good it will never need to change.'' 
Instead, we strive to organize and code the package such that enhancements
and maintenance changes will be as easy to incorporate as
possible.
  
In order to stand the test of time, we feel the minimization package
must include complete and easily understood user documentation, as well as
full mathematical documentation of all algorithms used.  

\subsection{Capabilities}

The key requirement is that every capability present in Minuit must be matched
in our minimization package.

It could be that Minuit is the ideal minimizer for HEP {\em per se}. 
Or perhaps the way physicists have grown accustomed to doing data analysis 
has evolved in tandem with Minuit to take advantage of its capabilities.
The point is moot--any package which cannot match the capabilities present in
Minuit will be unacceptable to the HEP community.

These capabilities lie in several areas.  
Clearly, the Minuit algorithms ({\sc migrad}, {\sc simplex}, {\sc minimize}) 
must be present.
And user code must be able to control steps taken and selection of
(and switching among) algorithms.  
Code must be the able to establish limits on some or all of the function
parameters, and fix and release their values.
And the methods for error analysis, nonparabolic error intervals,
and two-parameter contour lines (analogous to {\sc hesse}, 
{\sc minos}, and {\sc contour}) must be provided.  

It is tempting, along the way, to make any tweaks that are noticed to be pure
improvements.  
Some of these are pure extensions, such as the ability to define parameters
which have only an upper or lower limit.
Others might be minor algorithm improvements, such as a theoretically 
superior finite difference step size when numerically computing derivatives. 
In such cases, it is required that default behavior 
precisely mimic the procedures used by Minuit.

The Minuit-matching requirement is a matter of capability, and not of interface.  
Minuit assumes one of two usage paradigms:
\begin{enumerate}
\item
The minimizer is a static facility, which
the user code explicitly initializes to embark on a problem.
The user program then controls this facility by calling various
subroutines.
\item
A commanding minimizer program is supplied with ``cards,'' each directing
an aspect of initialization or an element of control.
\end{enumerate}
Clearly, both of these paradigms are obsolete in the context of a C++,
object-oriented package.  
The experienced Minuit user may miss the mantra of calling {\sc mnexcm}
and the
nine-argument calls to {\sc mnparm};
these will be replaced 
by methods which are more readable and which assigns one easily understood
purpose to each method called.
Still, when the user constructs a minimization problem and invokes problem
methods, she is logically employing capabilities which encompass
the same set as was available in Minuit. 
 
\subsection{Modern O-O Approach}

To make best use of the object oriented abilities provided by C++,
it is not enough merely to ensure that a minimization problem can
be used as an independant entity.  
We must use the tools at our disposal to make the setup and control of a
minimization problem as clear and flexible as possible.  
The focus is on user code readability,
and on
avoiding traps where the user can do something which looks
sensible but may have disasterous (or worse, subtly incorrect) consequences.

This means isolating conceptually distinct areas of control, and
providing ways of excercising control of one 
aspect of execution at a time.
Thus the user will be able to interact with termination conditions 
associated with a problem, and separately with the domain restricting
ranges of parameters, and with the selection of algorithms.

In each of these areas, the user controls behavior by interacting with 
specialized versions of objects.  
The minimization problem then interacts with these objects via
by their base class
interfaces.  
An example is in order:

To control how parameters are limited in range, the {\tt Problem} will own  
an instance of a {\tt Domain}.  
{\tt Domain} methods takes care of such necessities as translating 
{\tt Point}s
and {\tt Gradient}s between internal unrestricted Cartesian coordinates used by the
algorithm, and external parameters known to the function. 
But {\tt Domain} by itself is merely a base interface class; the actual 
{\tt Domain} attached to the {\tt Problem} would (by default) be a 
{\tt RectilinearDomain},
which provides the capabilities present in Minuit:  Restricting the
upper and/or lower limits on each parameter individually.
{\tt RectilinearDomain} inherits from {\tt Domain}.
The user who desires to alter these limits would obtain a pointer to 
the {\tt Domain}, and invoke methods of {\tt RectilinearDomain} such as 
{\tt setUpperLimit(n,x)}.  

Now suppose another user has a problem which
should be restricted to a region delimited by arbitrary planes in parameter
space.  This user would provide a different sort of object inheriting from
{\tt Domain}--say {\tt LinearConstraintsRegion}.  
And then she might call methods
of that class, such as {\tt forceLessThan(coefficients, maximum)}.  
The algorithm does not care which type of {\tt Domain} is in place.

In providing the user interface, it is tempting
to parallel the
control made available to Minuit.
This is good only insofar as there aren't opportunities to improve
on the semantics provided by the Minuit subroutines.
We are finding that there is a tendency to rely too heavily on 
doing what Minuit does, and we need to be careful to look for
opportunities to provide interfaces which would be clearer to people
who are not steeped in Minuit usage.  
To paraphrase Stroustrup, this package should be ``as close to Minuit
as possible, {\em but no closer}.''

\subsection{Extensibility and Maintainability}

The design of the package must be such that as new algorithms and 
anaylsis methods are developed, or as new types of termination conditions or
domains are required, they can be added either to the package itself, or
to one's local version of the package.  

We cannot insist that the work associated with such an enhancement
be small, because in principle coding some new algorithm might inherently
involve a lot of effort.  What we can, and do, insist on--and this drives
the entire package design--is that the additional work imposed by being 
part of this minimizaztion package is a very small fraction
of the typical work needed to create the new algorithm or domain itself.
This ``additional work'' includes 
\begin{itemize}
\item
Understanding all the steps one needs to do.
\item
Following the logic of existing package code if necessary.
\item
Discovering precisely where modifications will be needed, 
and what the nature of any new classes must be.
\item
Coding any C++ boilerplate needed to support the enhancement.
\end{itemize}
These are the tasks which represent a barrier to putting an enhancement 
into any of the various C++ Minuit translations, and we must ensure
that this barrier is small.

Moreover, we must see to it that the enhancer, who may well be an expert
in algorithm coding, need not also be an expert in C++.

The way to accomplish these goals is to carefully design the system such
that any two concepts that need not be logically intertwined, are expressed
as distinct classes which interact as little as possible.  The overall
package must be split into distinct subsystems, and dependencies carefully
controlled.
If this is done properly, then the enhancer need only be concerned about the 
small corner of the package this enhancement will involve.   
Thus all the code to be examined in creating 
a new algorithm class would be in the {\tt Algorithm} base class, 
and all the boilerplate involved would be found in classes in 
the Algorithm subsystem.

The same philosophy that leads to ease of enhancement without undue
tracing through an entanglement of code, will also greatly reduce
the burden of maintenance.
Over the long term for a non-commercial package, one has to assume that
any maintainer might well lose familiarity with the internals of
the system.   
So it is crucial that proper coding idioms be employed, to avoid setting
subtle traps.  And good separation of functionality
and control of dependencies are critical, so that when changes need to be
introduced, they will be localized and not have tendrils of effect throughout
the package.

Also, each class should have self-contained (and easy-to-automate) unit
tests; each subsystem should have integration tests only involving it and 
any subsystems it depends on; and a good suite of regression integration 
tests should be provided to help catch any errors introduced.

Having set a goal of easy enhancement, we must recognize that there are
different sorts of enhancements which will normally be performed by different
levels of developers.  A user might well want to create some new form of 
Termination condition; this must be very straightforward.  More rarely, 
a more experienced coder might want to derive a new type of {\tt Domain} (for
example, a {\tt RectilinearDomain}
but replacing the arcsin mapping function with
some sort of sigmoid).   Still more rarely,  
adding a fundamentally new algorithm is the sort of ehancement which
might be assisted by the principal package maintainers.

\subsection{Realm of Efficient Operation \label{efficiency}}

In the ``low-ceremony'' tight design iteration process so often sucessfully
used in High Energy Physics code development, certain fundamental design
considerations are often overlooked. 
Two such considerations are the 
required levels of agility and 
efficiency under various circumstances.  
In the course of designing this package, 
there will be opportunities for 
tradeoffs among areas such as coding style, 
levels of agility when faced with minor changes in needs, 
and performance efficiency.
An up-front grasp of the goals in these area will
allow the developer to make intelligent choices.  

We have discussed above the agility considerations, in terms of extensibility
and maintainability goals.  As to efficiency, the typical first reaction is 
to say ``of course this should be as efficient as possible,'' 
but that is an inappropriate goal for this package.
Taken to the extreme, a total focus on efficiency might dictate unmaintainable
coding tricks, avoidance of sound accepted C++ techniques, and ultimately
abandoning the object-oriented paradigm.
Some aspects of performance efficiency are critical, 
but the design goals should permit other forms of 
efficiency to be sacrificed in favor of better modularity, semantic clarity,
or other desirable features.

The HEP minimization problem has a natural set of performance goals, 
and these induce an efficiency philosophy which is implied in Minuit, and
which we state implicitly for our package:
It is assumed that the work involved in an invocation of the function being
minimized is very large compared to the overheads of 
 bookkeeping, 
 argument passing,
 and
 transformations of parameters.  
Thus tradeoffs which involve attaining beneficial properties
at the cost of incuring extra work in these bookkeeping areas
are considered positive.

Furthermore, it is also assumed that the number $N$ of parameters for
tractable minimization problems of the nature we are considering 
will not be too large.  
By this, we mean that even in an algorithm which requires operations 
(such as inversion) involving $N \times N$ matrices, 
the time spent for such operations does not dominate the time 
needed to evaluate the fuction in forming those matrices.

\subsubsection{Performance-based limitations}
 
This minimization package will perform well on problems in which the
dominant time cost is that of evaluating the function.
Because of the tradeoffs between some forms of efficiency and other 
desirable features, this  package may not perform efficiently 
for certain other classes of problems.  
It is tempting to say that such problems--which involve inexpensive
function calls--are so low-cost that nobody should care about performance,
so that this minimization package is suitable for all uses.   
We should realize
that the above rationalization is a bit facile; there are actually at
least four cases in which efficiency considerations may become important:
\begin{enumerate}
\item
Each call to evaluate the function is expensive.  
This is the case in most HEP fitting applications, 
and is the case which our package will deal with well.
\item
Some problems are potentially expensive not because each function call is
costly,
but because there are a large numbers of parameters, such that matrix
operations or even variable transformations begin to become dominant.
\item 
When the number of parameters is huge, the algorithm may require a huge
number of function calls.  If function calls are inexpensive relative to 
the number of parameters, then bookkeeping overhead can become significant.
The minimization problems appearing in Lattice QCD to find fermion propagators
are of this nature.
\item 
The overall scientific problem being solved may involve solving 
large numbers of easy minimization problems.  Thus, while no single minimization
is costly, performance on each problem might remain a critical consideration.
In that case, problem initialization costs can become important.
\end{enumerate}

The efficiency requirement laid on this minimization package,
or for that matter on Minuit, is that it is appropriate for efficient
solution of problems in the first of the above categories.
This has been found to meet the fitting and other minimization 
needs of most HEP applications other than Lattice QCD.

\section{C++ CONSIDERATIONS}

The immediate preferences of potential HEP users can come into tension
with the desired use of classical C++ 
coding idioms accepted by top practitioners.
Concepts which are legacies from 
the Fortran days lead physicsts to become comfortable with coding 
techniques which are today recognized as poor use of C++.  
In such cases, it is better to use the accepted modern idiom,
rather than to deliver inferior--but temporarily more familiar--interfaces.

Thus for example, it would be unwise to shun the use of 
{\tt auto\_ptr} when that is the appropriate way to convey the passing of
responsibility for an object.  
The assumption is made that physicists who may at first be uncomfortable
due to unfamiliarity with this idiom will quickly pick up on it.
And the benefits in using concepts which have been battle-tested by 
the world of developers, and which match the understandings of skilled 
programmers, are considerable.

Few of the decisions discussed in this section were black and white, 
and some may change in the course of coordinating with the CERN effort,
but after some consideration, the following principles were adopted:

\subsection{Function or Functor?}

How does the user supply the function to be minimized?  
The naive mechanism is that the user passes to the {\tt Problem}
a global-scope function of some specified signature (for example,
taking a vector of coordinates and returning a value).  In fact, this
notion is so obvious to the physicist coming from Fortran, that we
must support it in some manner.

But there are considerable advantages to being able to supply a functor
(a instance of a class that has an {\tt operator()} so that it looks like
it can be called as a function) instead.
One advantage to suppying a functor is that unlike the global function,
an instance of a functor may have internal state, and the class may provide 
methods to influence the function behavior.  (For example, one may wish to 
sum errors over fewer data points early in a minimization problem, and 
more later.)  

This natural concept in C++ replaces the awkward nature
of {\sc fcn} in Minuit, which contains six arguments.  
Only two arguments are fundamental to the notion of a function
to be minimized:  The vector 
{\sc xval} of coordinates, and the result {\sc fval}.
By allowing for a functor class, we obviate the need for 
{\sc iflag} and {\sc futil}; and by separating the gradient method
we eliminate the {\sc grad} argument.

\subsection{Object Ownerhip and Responsibility}

The package has several situations in which behavior is defined by hooking
into user code.
Two examples are the functor representing the function
to be minimized, and a {\tt Domain} object implementing limits on cooordinates.

The pattern of allowing a user to control behavior, 
by supplying a class which overrides one or more ``hook'' methods,
resonates well with physicist coders.  Were we to say that the user
constructs a {\tt Problem} supplying a reference or pointer to an instance of 
some class derived from {\tt Function}, 
the HEP community would not complain.  
And if we further warned the user not to allow this function to go out
of scope while the {\tt Problem} still needs it, most users would find that
an obvious caution. 

However, the consequence of violating that stricture are dire:  When
the {\tt Problem} calls (by pointer) the out-of-scope method, confusion will 
reign, with no sensible diagnostic to point out why.  Good library
developers protect against this by copying the vulnerable object.
The {\tt Problem}, then, will {\em own} the {\tt Function}
and {\tt Domain}, and not risk
having those objects ``pulled out from under it.''  
This notion is expressed by the
user passing an {\tt auto\_ptr} to a {\tt new} object (on the heap);
the {\tt Problem} assumes responsibility for deleting that object 
when it no longer needs it. 

Having made that decision, we must provide a way for the user to 
invoke controlling methods on the instance of the functor owned by the
{\tt Problem}.  
The original object that the user supplied to the {\tt Problem} won't do, 
since the {\tt Problem} will be using a copy of it; 
in fact, if the correct mantra
was used to pass ownership to the {\tt Problem}, the user code should 
no longer be able to 
access the original object.  Instead, the correct recipe is:
\begin{itemize}
\item
The user obtains a pointer to the base class (in this example, {\tt Function})
by calling a method of Problem (in this case, {\tt getFunction()}).
\item
This pointer is dynamic cast into the class the user needs to control.
The user can then invoke any control methods applicable for that class.
\item
The pointer is now allowed to go out of scope, without the user calling
delete on it.
\end{itemize}

The suggested syntax, which automatically obeys this recipe, involves using
the returned pointer 
as a temprary variable which never is valid outside a single
statement.  The syntax can look like:
\begin{verbatim} 
  dynamic_cast<RectilinearDomain>
    myProblem.getDomain()->
      setUpperLimit(paramNumber, 35.1);
\end{verbatim}

Because this mantra may not be immediately familiar to our users, 
it is important that the package 
stick to the same rules ({\em never} use delete on 
a pointer which Problem has supplied) for every case of object access. 

\subsection{Use of Templates}

The powerful template facility in C++ would allow clever ways to improve
the user interface and/or implementation of the Minimization package.
To give one example, methods giving a choice of supplying 
a user function or functor
to act as the minimization target
could be unified by expressing {\tt Problem} as a template,
taking the function class as a template parameter.

The use of templates has several adverse consequences which we perfer 
to avoid where practical:  It makes it a bit tougher for a user to 
comprehend a header file, it can make maintenance more difficult,
it makes compilation error messages much 
less readable, it introduces quirky linking issues, 
and it risks harming portability on certain weak compilers.
There are valid answers to each of these objections; still,
absent compelling benefits, 
the package design avoids the use of templates.

On the other hand, existing templated classes in the standard library are used
freely.

\subsection{Sophisticated Patterns}

Our philosophy concerning clever and subtle patterns to which convey
minor benefits is that it is sometimes possible to be {\em too} clever.

For example, the design of the {\tt Function} base class
(discussed in section \ref{function}) makes it possible
to supply a {\tt gradient()} method but 
forget to override {\tt gradientAvailable()}.
That is a bit of a trap--the algorithm would ignore the user-supplied
method when it needed a gradient.  (The same trap is present in Minuit, 
if the user forgets to call {\sc setgradient}).

We could finesse this by using a variant on the visitor pattern:  When 
the algorithm needs a gradient, it requests it from the {\tt Function} object 
but the signature of the {\tt gradient()} method 
also supplies the address of the
algorithm's finite difference gradient-computing method.  
The base class {\tt gradient()} invokes this method;
if the user's functor class overrides {\tt gradient()} 
that routine replaces the finite difference method behavior.
This pattern incurs some efficiency loss, but since that is merely a matter of 
an extra argument passed around, and an extra subroutine call, that cost would
(by the definition in section \ref{efficiency}) be acceptable.
However, the pattern looks complicated and mysterious; it might make 
maintenance more difficult and it might worry potential users who haven't
put in the time to understand it.  Since the gain is relatively slight,
we have come down on the side of the 
simpler {\tt gradientAvaliable()} solution.

\subsection{When to Introduce a Class}

In many cases, there may be multiple concepts for which the natural expression 
in bits and bytes will look like identical data structures.  For example,  
both the set of coordinates identifying a point, and the set of values 
representing the components of the gradient of a function, are underneath
indexed collections of doubles.  The question is, should the design introduce
distinct classes to represent the distinct concepts, or should it use a 
vector of doubles for both?

Good C++ design takes the view that if two concepts are distinct, they should
be represented by two distinct classes.  The idea is that you can't be sure in
advance that you will never alter the way you work with one of the concepts,
and if they are represented by distinct classes, making such a change will be
much less of a maintenance headache.
Introducing a distinct class to represent each 
distinct concept also provides some measure of type safety, 
in that you can't accidently use one concept when you mean the other.
And naming each concept leads to superior code readability.  
In a package that emphasizes extensibility, these three advantages 
are significant.
 
The disadvantage in introducing such classes is limited to the need to write
some boilerplate
code writing to provide their headers and near-trivial implementations. 
Therefore, especially concerning minimization package internals, 
we tend to introduce distinct classes for distinct concepts, 
particularly in the area of geometric concepts (see section \ref{geometric})
such as {\tt Point} and {\tt Gradient}.

\subsection{Namespace}

The minimization package must be a good C++ citizen.  By that, we mean it 
must not step on potential user symbols; we place everthing into a namespace 
to guard against this.  Other items we avoid are the use, in headers, of
macro definitions (which may clash with user defines) and of {\tt using}
statements (which defeat the purpose of the namespace protection if they
appear in an included header).

Namespace use has been approved for the CLHEP library, which has severe
portability requirements, so we are not concerned on that issue for the 
Minimization package. 

\section{LOGICAL DESIGN}

The package is divided into several subsystems.  A subsystem is 
defined as some group of classes and responsibilities, such that 
each subsystem depends on a minimal number of other subsystems, and such
that these dependencies can be expressed as cleanly as possible.
The first step in logical design was to identify the concepts present
in Minuit.  It is no coincidence that each subsystem coresponds to
one major concept in the minimization realm. 

The benefit 
of a good separation into subsystems comes during development
of various classes, and later during enhancement implementation, 
because the coder can have in mind a smaller ``overall picture''
when deciding how to do things.
Good isolation of subsystems makes it possible to develop the more 
detailed package design 
one part at a time, without constantly having to  consider the ramifications of
each decision on every other part of the design. 
Finally, the up-front thought about subsystem components has yielded 
immediate benefits in the form of solidifying the collection of concepts
involved in the minimization package.

In principle a ``subsystem'' need not be tied to a group of classes.
However,  
each identified subsystem in the Minimization package
is characterised by a key class embodying the 
interface of that subsystem to the user and/or to the other subsystems. 
This section will outline the responsibilities and dependancies of 
each subsystem, and discuss associated design issues.

\subsection{Problem}

The Problem subsystem has the responsibility for 
the user interface to  a minimization process.
It contains the {\tt Problem} class, which
takes ownership of {\tt Algorithm}, {\tt Domain}, and {\tt Termination}
objects, and contains the {\tt ProblemState}.

Minimizing involves the {\tt Problem} repeatedly requesting steps from the 
currently associated {\tt Algorithm}.  After each step, it determines whether
the user-supplied termination condition is satisfied, or whether the
algorithm itself has declared that it can't profitably proceed further.
If termination is reached, control returns to the user code, which may
change the {\tt Algorithm}, the {\tt Domain} 
(for example, releasing some parameters and
fixing others), and/or the Termination conditions, 
and again invoke the problem's
{\tt minimize()} method. 

\subsection{ProblemState}

{\tt ProblemState} is a structure containing information about the state
of solution which is not peculiar to any one algorithm or domain.
For example, the concept of the current best guess is inherent in
the nature of minimization; it is applicable whehter the algorithm 
producing it was {\tt migrad} or {\tt simplex}.

The user indirectly interacts with {\tt ProblemState}, via methods of 
{\tt Problem},
when she needs to inspect
this information.  
Subsystems of the minimization package interact with ProblemState
more directly.  For example, Termination conditions inspect its data
to decide whether to terminate, 
and the Algorithm subsystem is dependant upon {\tt ProblemState}, reading and 
altering its data. 

There is danger that {\tt ProblemState} can 
act as a catch-all for data that violates reasonable
encapsulation goals, effectively replicating--with all its flaws--the
notion of a big global common area. 
There is some art in restricting the contents to those items genuinely 
inherent in the minimization concept, and keeping those items which are
associated with an algorithm in the specific Algorithm classes.
 
\subsection{Algorithm}

The {\tt Algorithm} base class provides interface between the {\tt Problem} and 
subclasses implementing specific minimization algorithms.  This base class
also standardizes ways for its derived classes to access the function and
domain objects.  For example, the mantra for obtaining a function gradient
involves finding out whether the {\tt Function} 
object can provide it analytically,
and if not, utilizing the algorithm's own finite difference methodology;
this mantra is encapsulated in {\tt Algorithm}.

Any given algorithm class has an {\tt iterate()} method, 
which ``takes a step,''
improving the state of solution.  
It also decides whether to declare that this algorithm is finished
in the sense that no further meaningful improvement is forseen.
Between steps, the {\tt Problem} applies user-supplied {\tt Termination}
conditions
which may declare victory well before the algorithm declares exhaustion.

This subsystem depends on the Function and Domain subsystems:
Algorithms, of course, repeatedly requests values of the function (and if
available its gradient and second derivatives), and utilize the {\tt Domain} to 
map its internal coordinates into external parameters to supply to those 
function calls.

\subsection{Analysis}

This subsystem has the responsibility for the various solution analysis
tools such as {\sc minos}, {\sc hesse}, and {\sc contour}.
Such tools have roughly the same needs and dependencies as algorithms, 
but a class derived from {\tt Analysis} can in addition  
also depend on Algorithms which it needs to call.

\subsection{Function \label{function}}

The Function subsystem has the responsibility for providing function values
and, if available, analytic  
gradient and second derivatives.
The primary class is the abstract interface {\tt Function}.  

Commonly, users will have an ordinary function
(taking a constant refernce to a {\tt Point} or to a vector of doubles,
and returning a double) to minimize.
Such users can make use of the free method {\tt UserFunction()}, 
to supply this function when constructing the {\tt Problem}, as in:
\begin{verbatim}
  double f(const vector<double> v) {
    assert (v.size()=nargs; 
    // ... do work to return a value }
  Problem myprob ( UserFunction (nargs, f) );
\end{verbatim}
\noindent
An alternative signature would take both the function and a method to
compute the gradient, if direct gradient computation is available.
The user can also optionally supply 
global methods to compute second derivatives.

{\tt UserFunction()} constructs a {\tt Function} object on the heap, 
and returns an auto\_ptr to it;
in the example code, this is immediately passed to the constructor of the 
{\tt Problem}.

Alternatively, a user
can inherit from {\tt Function} to form his own ``functor'' class, and supply 
an instance of that when constructing a {\tt Problem}. 
In that case the functor class
must override the method 
which returns the function value at a point.  
It may also override the {\tt gradient()} method, and if it does, 
it should override {\tt gradientAvailable()} to return {\tt true}--this
replaces Minuit's {\sc setgradient} command, and more importantly, 
obviates the need for the
user function to be passed an {\sc iflag} argument 
to tell it what do do.
Similarly, the functor
class may override the {\tt hessian()} and {\tt hessianAvailable()} methods.

The functor is to be created on the heap and
an {\tt auto\_ptr} supplied to the {\tt Problem} 
so that there is no danger of the   
user getting mysterious errors due to the functor going out of scope
while the {\tt Problem} is depending on it.  
The user can later recover a pointer to this function object and thus
invoke its methods:
\begin{verbatim}
  class MyFunct : public Function {
    MyFunct() : Function(nargs){}
    double operator()(const Point& p);
    changeStuff(int c);
    ...};
  Problem myprob ( auto_ptr<Function>
                  (new MyFunct)       );
  (dynamic_cast<MyFunct*>
     (myprob.getFunction()))->changeStuff(3);	  
\end{verbatim}

The Function subsystem depends only on the Geometric Concepts subsystem.

\subsection{Domain}

The Domain subsystem has the responsibility for providing mappings between
the internal coordinates that an algorithm works with, and the external
parameters that a function deals with.
This solidifies the decoupling, already present in Minuit,
between the algorithm and any restrictions on the parameter values.
That is, it would be too difficult to invent an algorithm that could
deal with any arbitrary sort of variable restrictions; instead, the 
algorithms themselves work with clean, unrestricted Cartesian coordinates,
and all the complexity involving restrictions is separated off into the 
Domain classes.  The {\tt Domain} base class provides the interface for
algorithms and other parts of the package to transform variables.

There is, of course, the complication that 
geometric objects of different
kinds transform differently under the mappings provided by a Domain.
Thus specific {\tt Domain} derived classes provide methods 
to map various Geometric Concepts: {\tt Point}s, {\tt Gradient}s, 
{\tt Hessian}s, 
{\tt CharacteristicScale}s, (error) {\tt Correlations},
and whatever else algorithms and analyses may need.
The base {\tt Domain} class 
supplies these methods in terms of virtual functions 
which supply the map's value and derivatives.
 
Domain encapsulates the Minuit concepts of parameter limits and of
fixing/releasing parameter values.  
Half the interface of Minuit is devoted to manipulating
the way parameters can vary.  Most of this control belongs in the Domain
subsystem.

The exemplar of a {\tt Domain} 
(and the one associated with a {\tt Problem} by default)
is the {\tt RectilinearDomain}, 
which provides the mapping functions and capabilities
present in Minuit.  A user could fairly easily create a class inheriting
from {\tt RectilinearDomain} in order to modify the mapping functions used.

{\tt RectilinearDomain} has the property that its individual components map in
a separable way, that is, the map of each internal component depends on
only one variable external parameter.
Algorithms can query a {\tt Domain} about this property.  
For example, when {\sc migrad} 
utilizes diagonal second derivatives to avoid the need for computing the
full Hessian, it relies on this separablity property; and in more complex
Domains, additional work may be needed.

Unlike functors derived from {\tt Function}, 
which need only supply the fundamental
value-suppying method (and may optionally supply gradient etc.),
each {\tt Domain} subclass is required to supply all the Geometric Concepts
mapping methods.  
While typical users are expected to create functors deriving from 
{\tt Function}, 
we can assume that those who embark on defining new types of Domains
are somehwhat more expert.

The Domain subsystem depends only on the Geometric Concepts subsystem. 

\subsection{Termination}

The Termination subsystem has the responsibility for conditions
conditions which the 
problem can apply, to
determine whether to terminate the minimization iteration 
and return control to the user.
The {\tt Termination} base class also defines an interface for potential 
user-created termination conditions.

The subsystem contains Termination classes corresponding to each 
of the criteria applicable in Minuit, plus a time-based criterion. 
Time-based termination is often preferable to a count of function calls,
but has a potential disadvantage with respect to portability.  
However, the standard library does provide for timing at the 
one-second granulaty level, and almost all systems are compliant in this
respect.

Termination also comes with methods to build ``compound'' Termination objects
as by logically combining ({\sc and, or}) more basic Termination objects.
Such a compound criterion could, without  
too much difficulty, be
coded by each user as a class deriving from {\tt Termination}.
But the desire to apply a group of criteria, terminating (say) if 
either the accuracy criterion is met, {\em or} the number of function calls
is excessive, will be a very common case.  
So it was decided to make life easy for such users by 
supporting these combinations.  
We supply the logical combiners {\sc and} and {\sc or},
as this will satisfy the vast majority of applications.

The Termination subsystem depends on ProblemState.

\subsection{Geometric Concepts \label{geometric}}

The minimization process deals with a number of geometric
concepts, which are 
logically distinct and not interchangeable.
Most of these behave like a vector of doubles.  
For example, a Point in exterior coordinate
space is not interchangeable with a Gradient of the function at some 
point, even though they both have the same number of components.  And
neither is the same as an ``InteriorPoint'' which is a set of 
coordinates in the Cartesian space in which the algorithm is working.

Although the package code utilizes these Geometric Concept classes,
there is good reason to provide user interfaces for common operations
which deal with the familiar vector of doubles.  For example, the 
{\tt UserFunction()} method 
can accept a function of a {\tt std::vector} where a {\tt Point} 
is logically called for:
\begin{verbatim}
  double f (const std::vector<double> params);
  Problem myprob ( UserFunction (nargs, f) );
\end{verbatim}

The exterior concepts identified are {\tt Point}, {\tt Gradient},
{\tt SecondDerivatives}, {\tt Hessian}, {\tt CharacteristicScale}, and 
{\tt Correlations}.  There are also classes corresponding to each of these,
representing internal-coordinate concepts.

The Geometric Concepts subsystem contains the collection of such classes.
It depends on no other subsystem.

\section{IMMEDIATE ENHANCEMENTS}

The initial release of the minimization 
package is, by definition, to include all Minuit 
functionality.
Of course, the object nature of a minimization problem will allow
modes of usage which would be awkward with the Fortran Minuit, 
but basically the capabilities will at that point be those of Minuit.
However, certain extensions will be in the initial release.
These appear in situations where the nature of a potential
extension is obvious, and the implementation is straightforward.
If the extension is clear in definition, and 
we have a firm idea of how it should by implemented,
then there is no reason to delay that capability unitl a later release.

In the Domain subsystem, the {\tt RectilinearDomain} 
class will immediately support
half-open intervals for parameter limits (that is, 
having only an upper or lower
limit for a paramenter).  It will also be easy to substitute for the arcsin
mapping functions.

In the Function subsystem, the package will allow supplying
second derivatives (either diagonal or full matrix) along with
the gradient.

In the Termination subsystem, several new basic choices will be provided,
including a time-based termination, and arbitrary logical combinations
of termination conditions will be supported.

\section{STATUS}

A preliminary partial implementation of this package has been done,
to obtain familiarity with the Minuit algorithms and concepts, and 
to get a feel for the design issues.  (A spin-off of this work is 
a set of heavily-commented Minuit Fortran ``source'' code, 
and some mathematical documentation of the Minuit algorithms.)   

The package design is completed (though not set in stone) and most
interfaces have been defined and headers coded.  
Re-implementation of the classes, based
on the actual design, is in progress.

The CERN effort (as discussed with F. James) 
is at the point where the minimization algorithms (but not all
of the analysis algorithms) have first-pass implementions, again in a
familiarization phase.  Since the two efforts are at comparable stages, 
the prospects for coordination seem reasonable.

\begin{acknowledgments}
The authors wish to acknowledge substantial assistance in package design  
and C++ technique from Walter Brown and Marc Paterno.

This work was performed at Fermilab, which is operated by the Universities
Research Association, under contract number DE-AC02-76CH03000 with the U.S.
Department of Energy.
\end{acknowledgments}


\end{document}